\documentclass[preprint,aps]{revtex4}
\usepackage{graphicx}
\begin{document}
\title{Connectivity and Critical Currents in Polycrystalline MgB$_2$}

\author{M Eisterer, J Emhofer, S Sorta, M Zehetmayer, H W Weber}

\address{Atominstitut der \"Osterreichischen Universit\"aten, Vienna University of Technology, 1020
Vienna, Austria}


\begin{abstract}
Current transport in polycrystalline magnesium diboride is highly
non-uniform (percolative) due to the presence of secondary phases
and also due to the intrinsic anisotropy of the material. The
influence of secondary phases on the transport properties of
MgB$_2$ was investigated. Bulk samples were prepared from a
mixture of MgB$_2$ and MgO powders by the ex-situ technique in
order to vary the MgO content systematically. The samples were
characterized by resistive and magnetization measurements. The
reduced MgB$_2$ fraction is modeled by a reduced effective cross
section (connectivity), which was assessed directly by the
experiments. The presence of MgO also increases the percolation
threshold, which reduces the zero resistivity (or irreversibility)
field.

\end{abstract}

\maketitle

\newpage

\section{Introduction}

Magnesium diboride is an interesting material for applications. It
is cheap and can potentially be operated at higher temperatures
than the conventional superconductors NbTi or Nb$_3$Sn.
Unfortunately, the upper critical field, $B_{c2}$, of  pure
MgB$_2$ is comparatively small ($\sim 14$\,T at 0\,K
\cite{Eis07rev}) and the upper critical field anisotropy
\cite{Eis07rev} leads to a strong magnetic field dependence of the
critical current density, $J_\mathrm{c}$ \cite{Eis03}. Thus the
in-field performance of clean MgB$_2$ is modest, but can be
improved by the introduction of small impurities, which act as
scattering centers, reduce the mean free path of the charge
carriers and, therefore, enhance $B_{c2}$. Today's best wires
already touch the performance of NbTi \cite{Her07,Hur08}. It was
shown in Ref. \cite{Eis07rev}, that the critical currents in
MgB$_2$ are still smaller by a factor of about five compared to
expectations for an optimized material. It is non trivial to
decide whether this difference is caused by comparatively weak
pinning (compared to optimum) or by a reduced cross section over
which the currents flow \cite{Row03}, but it seems that the latter
is more important \cite{Eis07rev}. In fact, the density of the
superconducting filaments in typical wires or tapes is often only
around half the theoretical one and secondary phases, especially
oxides \cite{Bir08,Jia06,Kli01} or boron rich compounds
\cite{Kli01}, are found at grain boundaries. The latter are
expected to block the current flow. The corresponding reduction in
cross section is usually called reduced area or connectivity
problem and can be quantified by a factor $A_\mathrm{con} \leq 1$.
The current has to meander between the well connected grains and
this current percolation is amplified by the upper critical field
anisotropy of MgB$_2$, when a magnetic field is applied. The
grains of the current carrying ``backbone'' attain different
properties according to their crystallographic orientation with
respect to the applied field. A model for the current flow in
MgB$_2$ was proposed in Ref. \cite{Eis03}. It is based on the
introduction of an effective cross section $A_p\sigma_0$ (similar
to $A_\mathrm{con}$) as a function of the fraction of
superconducting grains, $p$, which changes with field,
temperature, and applied current. $\sigma_0$ denotes the
geometrical cross section, thus $A_p$ varies between 0 and 1. The
original model assumed in principle that all grains (or sites in
the terminology of percolation theory) consist of MgB$_2$ and are
perfectly connected. If these assumptions become invalid,
adoptions have to be made, which depend on the actual defect
structure, as discussed in the following paragraphs.

Each non-superconducting inclusion (or pore) reduces
$A_\mathrm{con}$, which becomes an additional parameter of the
model \cite{Eis07rev}. The percolation threshold $p_\mathrm{c}$
(defined as the minimum fraction of superconducting grains which
is necessary for a continuous current path) is expected to change
only if the average number of connections between neighbouring
(superconducting) grains is reduced. The number of connections
between nearest neighbors is constant in a regular lattice and
called coordination number, $K$. The percolation threshold is
closely related to the coordination number \cite{Mar97,Wie05}, the
larger $K$, the smaller is $p_\mathrm{c}$.

An instructive and simple situation occurs, if the morphology of
the defects is comparable to the morphology of the MgB$_2$ grains.
The structure of the system does not change, but the number of
superconducting grains is reduced to $p_{\mathrm{MgB}_2}p$.
$p_{\mathrm{MgB}_2}$ is the fraction of MgB$_2$ grains and $p$
denotes further on the superconducting fraction among the MgB$_2$
grains (which depends on temperature, field, and current). If
$A_p$ is a simple power law of the form
\begin{equation}
A_p(p,p_c)=\left(\frac{p-p_c}{1-p_c}\right)^t, \label{Ap}
\end{equation}
it can be rewritten as \cite{Eis07c}
\begin{eqnarray}
A_p(p_\mathrm{MgB_2}p,p_c) & = &
\left(\frac{p_\mathrm{MgB_2}p-p_c}{1-p_c}\right)^t \nonumber \\ &
= &
\left(\frac{p_\mathrm{MgB_2}(p-p_c/p_\mathrm{MgB_2})}{1-p_c}\right)^t
\nonumber \\ & = &
\left(\frac{p_\mathrm{MgB_2}(p-p_c^*)}{1-p_c^*}\frac{1-p_c^*}{1-p_c}\right)^t
\nonumber
\\ & = &
\left(\frac{p-p_c^*}{1-p_c^*}\frac{p_\mathrm{MgB_2}(1-p_c^*)}{1-p_c}\right)^t
\nonumber \\& = &
\left(\frac{p-p_c^*}{1-p_c^*}\right)^t\left(\frac{p_\mathrm{MgB_2}-p_c}{1-p_c}\right)^t
\nonumber \\& = & A_p(p,p_c^*)A_p(p_\mathrm{MgB_2},p_c) \nonumber
\\ & = & A_p(p,p_c^*)A_\mathrm{con}. \label{Apred}
\end{eqnarray}
Thus, $A_p(p_\mathrm{MgB_2},p_c)$ splits into a field and
temperature independent factor, which is just the connectivity,
$A_\mathrm{con}$, and a modified $A_p$ needed for the percolation
model \cite{Eis03}.  $A_\mathrm{con}$ is given by the same power
law as the original $A_p(p,p_c)$, with $p$ equal to
$p_{\mathrm{MgB}_2}$. The percolation threshold in the modified
$A_p$ increases to $p_c^*=p_c/p_{\mathrm{MgB}_2}$. This increase
reduces the zero resistivity field and increases the field
dependence of the critical currents \cite{Eis05}. Note that the
increase of the percolation threshold is independent of the actual
function of $A_p$, since it just results from the condition that
the total fraction of superconducting grains, which is reduced by
$p_{\mathrm{MgB}_2}$, still has to be $p_c$.

The same scenario occurs if individual MgB$_2$ grains are
completely covered by an insulating layer and disconnected from
the connected matrix, $p_{\mathrm{MgB}_2}$ then refers to the
fraction of connected MgB$_2$ grains.

If the insulating inclusions (the same holds for pores) and their
separation are much larger than the superconducting grains, mainly
$A_\mathrm{con}$ is reduced, since the average coordination number
does not change significantly. Only grains neighboring these
non-superconducting sites, have less connections to MgB$_2$
grains, but their number is small, since the distance between the
inclusions is assumed to be much larger than the diameter of the
MgB$_2$ grains. The reduction in $A_\mathrm{con}$ is expected to
have a similar dependence on the non superconducting volume
fraction for both small and large inclusions.

If the inclusions are much smaller than the MgB$_2$ grains, or if
they have a totally different morphology, it is hardest to predict
their influence. The blockage of  individual connections (not
whole grains) by insulating phases is expected to result in an
increase of the percolation threshold (reduction of the average
coordination number) and a reduction in $A_\mathrm{con}$. Since
the insulating layers might be much thinner than the diameter of
the MgB$_2$ grains, a small volume fraction of secondary phases
might result in a strong effect. The problem changes from site to
bond percolation \cite{Sta92} and the volume fraction is not the
ideal parameter in this case. In principle one expects a similar
dependence of $A_\mathrm{con}$ on the fraction of conducting
bonds, although with a different percolation threshold. (It
becomes a mixed bond/site percolation problem, when a magnetic
field is applied). However, the fraction of conducting bonds is
difficult to asses in real systems, but $p_c$ of the site
percolation problem might increase, because of a reduction in the
average coordination number of the MgB$_2$ matrix.

Small inclusions (compared to the MgB$_2$ grain size) are not
expected to influence $p_\mathrm{c}$, if they do not block whole
grain boundaries, but only reduce the conducting area between the
grains, or, if they are incorporated into the MgB$_2$ grains. The
reduction in $A_\mathrm{con}$ might be (over-)compensated by the
potential pinning capability of small defects. If these small
inclusions block whole connections, because they cluster at
individual grain boundaries, their influence should be similar to
insulating layers.

Although the reduction of $J_c$ by poor connectivity is quite
obvious \cite{Row03,Row03b}, only a few studies were aiming at a
quantitative relationship between the amount of secondary phases
or pores and the corresponding reduction of the critical currents
so far \cite{Gri06,Yam07,Mat08}. In this work, the
non-superconducting volume fraction is changed by MgO particles,
which were added to the MgB$_2$ powders before sintering. The
influence of the decreasing MgB$_2$ volume fraction on the
resistivity and on the superconducting properties is investigated
and described by changes in $p_\mathrm{c}$ and $A_p$.

\section{Experimental\label{secexp}}

Powders of MgB$_2$ (Alfa Aesar, 98\,\%, -325\,Mesh) and MgO (Alfa
Aesar, 99.95\,\%, -325\,Mesh), were weighed, mixed, and pressed
uniaxially with 750\,MPa. The nominal MgO content ranged from 0 to
60\,vol\%. The pressed pellets had a diameter of 13\, mm and a
thickness of around 2.6\,mm. Each sample was placed into an iron
container and sintered separately in argon atmosphere for 10\,h at
1035\,$^\circ$C. The temperature was ramped up with
500\,$^\circ$C/h. The sample cooled down slowly after the heat
treatment (within a few hours). The iron container was necessary
to prevent magnesium loss, which was monitored by comparing the
weight of the pellet prior to and after sintering. A stripe
(typically $10\times 1.6\times 1.5$\,mm$^3$) was cut from the
central part of the sample for the resistive measurements. A slice
(around $2\times 1.5\times 0.4$\,mm$^3$) was prepared from the
central part of the stripe after the resistive measurements for
the assessment of the critical current density.

The nominal volume fraction of MgB$_2$ was calculated from the
weight and the theoretical density of the mixed powders. Since the
density of the samples was only around 70\,\% of the theoretical
density, this nominal volume fraction strongly overestimates the
real volume fraction. Therefore, this quantity is denoted as
nominal content in the following. A better estimate of the real
volume fraction was calculated from the weight of the admixed
MgB$_2$ powder, the theoretical density of MgB$_2$ and the final
volume of the sintered pellet. (It is always close to the nominal
MgB$_2$ content times 0.7). The real volume fraction of MgB$_2$
might be smaller, if MgB$_2$ partly decomposes or reacts with MgO
during sintering. However, the volume fraction used in the
following always refers to this calculation.

The temperature dependence of the ac-susceptibility was measured
in a commercial SQUID magnetometer (Quantum Design), with an
amplitude of 30\,$\mu$T. The demagnetization factor of each sample
geometry (typical dimensions: $2\times 2\times 1.5$\,mm$^3$) was
calculated numerically (between 0.45 and 0.65) in order to
determine the susceptibility correctly.

The resistivity was monitored at various fixed applied fields
between 0 and 15\,T, while ramping the temperature down at
10\,K/h. Current and voltage contacts were made with silver paste.
An electric field of around 15\,$\mu$V/cm (max 25, min
5\,$\mu$V/cm) was applied at room temperature to ensure that the
(local) current density within the current path remains comparable
for all MgO contents. This procedure is based on the assumption
that the intrinsic resistivity of the conducting material
(MgB$_2$), $\rho$, does not change and that the increase in
resistance results only from the reduction in cross section of the
actual current path. The (local) current density $J$ is then only
given by the electric field $E$
($\overrightarrow{E}=\rho\overrightarrow{J}$). The current was
then kept constant (100\,nA to 10mA) during the measurement. The
corresponding (local) current density can be estimated to be
around $7\times 10^3$\,A\,m$^{-2}$. A linear fit to $\rho(T)$ was
made in the range $0.5\rho_n<\rho<0.9\rho_n$. Extrapolation to the
normal state resistivity $\rho_n(T)$ defined the (onset)
transition temperature $T_c$. $\rho_n(T)$ was also obtained by
linear extrapolation of the normal state resistivity above the
transition to lower temperatures. The temperature, where the
resistivity apparently reduces to zero (zero resistivity
temperature), $T_\mathrm{\rho=0}$, was evaluated. The upper
critical field and the zero resistivity field were obtained by
inversion of $T_c(B)$ and $T_\mathrm{\rho=0}(B)$, respectively.

Magnetization loops were recorded in a commercial Vibrating Sample
Magnetometer (VSM) at 5 and 20\,K. The field was ramped with
0.25\,T/min from -2\,T to 5\,T and back to zero. The critical
current density was calculated from the irreversible component of
the magnetic moment using the Bean model \cite{Bea64}. A self
field correction was made numerically \cite{Wie92}.

\begin{figure} \centering \includegraphics[clip,width=0.5\textwidth]{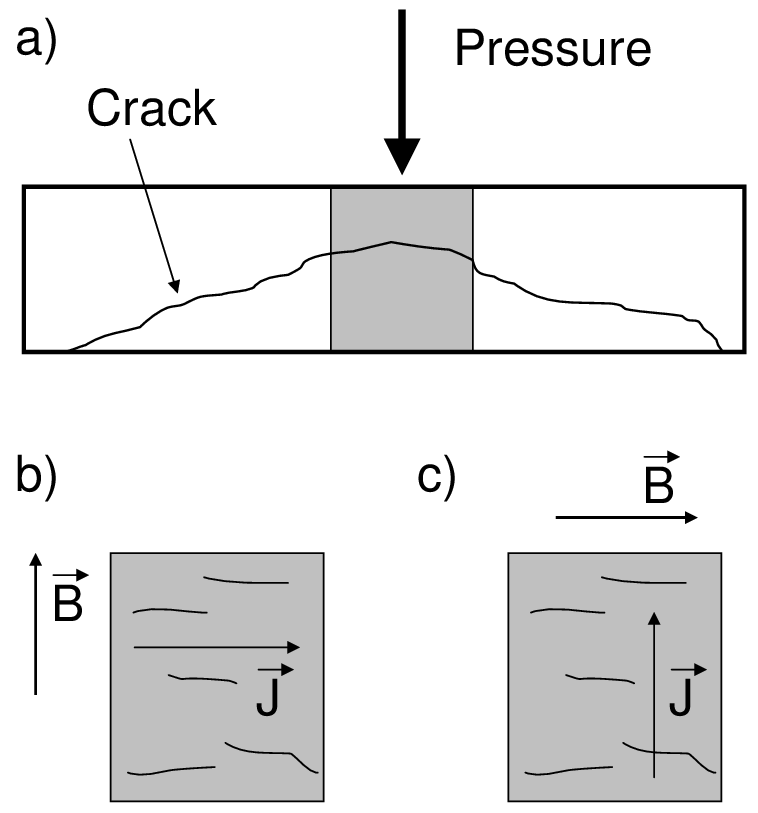}
\caption{Large crack in the pellet after pressing (a). Samples for
measurements were cut from the center of the pellet (gray area).
Orientation of the small cracks, the magnetic field and the
currents during resistive (b) and magnetization (c) measurements}
\label{Figgeo}
\end{figure}

A small anisotropy of the properties was detected in the
magnetization measurements. The zero resistivity field is lower by
about 3\,\% in the pure MgB$_2$ sample when the field is oriented
parallel to the direction of the uniaxial pressure, than after
rotation by 90\,$^\circ$. This means that the induced texture is
very weak and can be neglected in the following. The self field
critical current density is higher by about 75\% in the parallel
orientation. This significant difference is possibly caused by
small cracks. Large cracks were visible in some pellets, which
were perpendicular to the direction of the pressure (at least in
the center of the pellet, from where the samples were cut). This
is illustrated in Fig.~\ref{Figgeo}a. No samples with large
visible cracks were used in the measurements, but it is not
unlikely that small cracks were present in all samples, which
reduce currents flowing perpendicular to the crack direction
(Fig.~\ref{Figgeo}c). Only in magnetic fields parallel to the
former direction of the pressure, the induced shielding currents
always flow parallel to the cracks, which does not result in a
suppression of $J_\mathrm{c}$ (Fig.~\ref{Figgeo}b).

The magnetization measurements of all other samples were performed
only for one orientation, i.e. with the smallest dimension
parallel to the applied field. The field was always perpendicular
to the direction of the pressure (Fig.~\ref{Figgeo}c). In order to
check for possible influences of the geometry of the specimen on
the derived current densities, the pure MgB$_2$ sample was also
measured with the smallest dimension (and the direction of
pressure, which was the longest dimension) perpendicular to the
applied field. Nearly identical results were obtained in this
case.

The current was applied perpendicular and the field parallel to
the direction of the pressure during the resistivity measurements
(Fig.~\ref{Figgeo}b).

\section{Results and Discussion\label{secresults}}

\begin{figure} \centering \includegraphics[clip,width=0.5\textwidth]{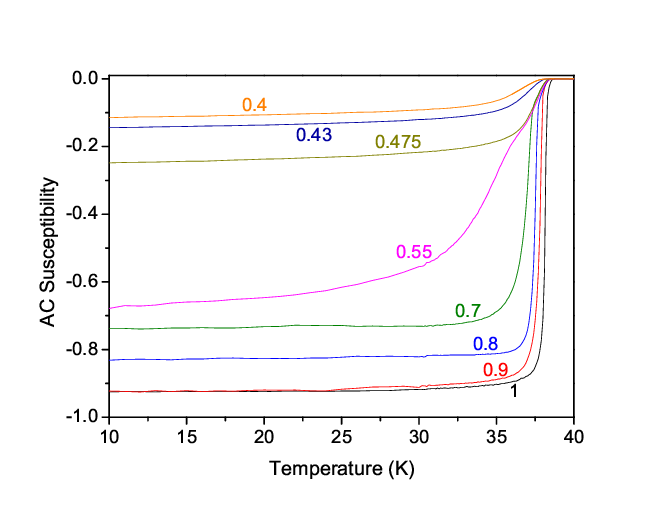}
\caption{Change of the ac susceptibility with the nominal MgB$_2$
content, which is given by the labels.} \label{Figsuscept}
\end{figure}

The ac susceptibility of samples with different MgB$_2$ content is
plotted in Fig.~\ref{Figsuscept}. The labels refer to the nominal
MgB$_2$ content. The samples containing 0 and 10\,vol\% MgO nearly
(-0.925) reach the ideal value, -1, for complete flux expulsion at
low temperatures. The slightly smaller value could result from an
overestimation of each dimension by only 2.5\,\%, or by the
imperfect geometry of the sample (the numerical calculation of the
demagnetization factor is based on a perfect cuboid). The low
temperature susceptibility then decreases continuously with
MgB$_2$ content, down to 55\,\% MgB$_2$. This is the first sample
with a quite significant broadening of the transition and a non
constant susceptibility below 25\,K. The further decrease in
MgB$_2$ content led to an abrupt reduction of the susceptibility.
The samples with 55 and 47.5\,\% MgB$_2$ behave quite similarly
above 37\,K, but the susceptibility differs by a factor of about
2.7 at low temperatures. Thus, macroscopic shielding was
established in the 55\,\% sample at low temperatures, while for
lower MgB$_2$ contents the length scale of shielding obviously
changes. Note that all corresponding transport samples were
macroscopically superconducting (zero resistivity), only the
40\,\% MgB$_2$ sample was apparently insulating.

The superconducting transition temperature, $T_\mathrm{c}$,
continuously decreases with MgB$_2$ content, although the maximum
shift is only about 0.5\,K near the onset of the transition (first
points deviating from zero). A larger reduction of $T_\mathrm{c}$
was observed at the onset of the resistive transition: from 38.7 K
to 37.8 K, for 100\,\% and 46\,\% MgB$_2$, respectively. Zero
resistivity was observed at 38\,K and 34.6\,K. The residual
resistivity ratio decreased from 3.7 in the pure to 1.9 in the
46\,\% sample and the normalized resistivity
$\rho_\mathrm{norm}:=\rho(40\,\mathrm{K})/\Delta\rho$ increased by
a factor of 3. The phonon contribution to the resistivity,
$\Delta\rho:=\rho(300\,\mathrm{K})-\rho(40\,\mathrm{K})$, is
expected to change hardly with disorder \cite{Eis07rev}, thus the
increase in $\rho_\mathrm{norm}$ reflects the enhancement of the
residual resistivity, $\rho_0$ ($\sim\rho(40\,\mathrm{K})$ in the
present samples). A linear dependence of $T_\mathrm{c}$ on
$\rho_\mathrm{norm}$ is observed, similar as in previous reports
\cite{Eis07,Eis07rev,Gan05,Tar06}. The corresponding increase in
upper critical field is observed in the present samples, too. The
changes indicate that the intrinsic properties of the MgB$_2$
grains changed somewhat, either by a slight loss of magnesium, or
by any reaction between the two different powders. In any case
impurity scattering is increased, which complicates the situation,
since not only the volume fraction of MgB$_2$ is varied. In
particular the upper critical field anisotropy, $\gamma$, is
needed in the following. It is known to be sensitive to impurity
scattering \cite{Kru07}, but it can be estimated from
$T_\mathrm{c}$ by \cite{Eis07rev}
\begin{equation}
\gamma(T_\mathrm{c})=\frac{t_\mathrm{c}^2+16.7t_\mathrm{c}(1-t_\mathrm{c})}{3.88-3.724t_\mathrm{c}},
\label{gammaTc}
\end{equation}
with $t_c:=T_\mathrm{c}/T_\mathrm{c0}$ and
$T_\mathrm{c0}=39.43$\,K.

\begin{figure} \centering \includegraphics[clip,width=0.5\textwidth]{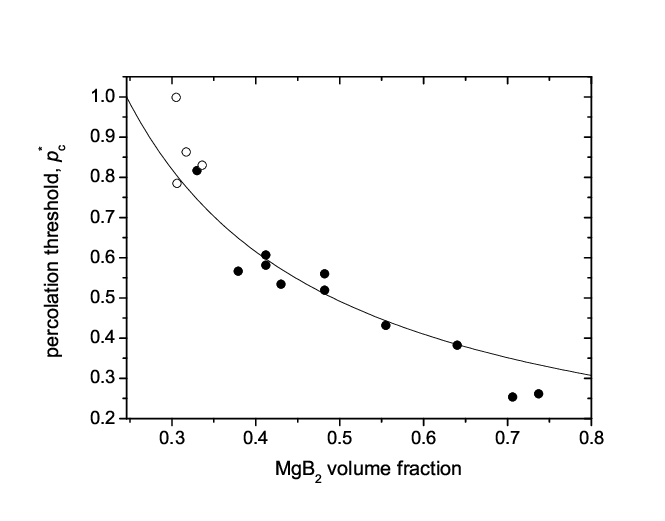}
\caption{The percolation threshold as a function of the volume
fraction of MgB$_2$, $p_\mathrm{MgB_2}$. The open symbols
represent data from samples with finite resistivity in the
superconducting state.} \label{Figpc}
\end{figure}

The percolation cross section $p_\mathrm{c}^*$ can be estimated
from the ratio between the upper critical field and the zero
resistivity field, if $\gamma$ is known \cite{Eis05}
\begin{equation}
p_\mathrm{c}^*=\sqrt{\frac{\frac{B_\mathrm{c2}^2}{B_\mathrm{\rho=0}^2}-1}{\gamma^2-1}}
\label{pc*}
\end{equation}
For highly diluted MgB$_2$, when $p_\mathrm{c}^*$ is close to 1,
$\gamma$ approaches $B_\mathrm{c2}/B_\mathrm{\rho=0}$. The highest
value for this ratio was 4.75, as found in the sample with the
highest, but still finite resistivity (44\,\% MgB$_2$). Exactly
the same value for the anisotropy is obtained from
Equ.~\ref{gammaTc} with the midpoint $T_\mathrm{c}$ (defined as
the temperature, where the resistivity drops to 0.5$\rho_0$). The
midpoint $T_\mathrm{c}$ is a reasonable criterion, since it
accounts for material inhomogeneities. In this particular sample
it was lower by 0.8\,K than the onset $T_\mathrm{c}$. This
excellent agreement justifies the application of
Equ.~\ref{gammaTc} to estimate $\gamma$ of all samples from the
midpoint $T_\mathrm{c}$. The percolation threshold
$p_\mathrm{c}^*$ was then calculated from the ratio of the upper
critical field and the zero resistivity field (Equ.~\ref{pc*}) at
15 K. This temperature was chosen to avoid extrapolation of
$B_\mathrm{c2}(T)$ to lower temperatures and $\gamma$ was shown
not to change significantly at lower temperatures.

Fig.~\ref{Figpc} presents $p_\mathrm{c}^*$ as a function of the
MgB$_2$ volume fraction. The line graph represents the theoretical
behaviour $p_\mathrm{c}^*=p_\mathrm{c}/p_{\mathrm{MgB}_2}$ with
$p_\mathrm{c}=0.246$ as obtained by a fit to the solid circles.
This value corresponds formally to the percolation threshold of
dense pure MgB$_2$ ($p_\mathrm{MgB_2}=1$). The open circles were
neglected, since they represent data from samples with a constant
but finite resistivity below the superconducting transition in
zero field. This indicates either the presence of conducting but
not superconducting phases in the remaining highly percolative
current path, or normal conducting tunnel currents (see below).
With these data points, $p_\mathrm{c}$ changes to 0.259. In the
following we assume $p_\mathrm{c}=0.25$, which is in between the
percolation threshold of a simple cubic lattice ($K=6$,
$p_\mathrm{c}=0.311$) and a face-centered cubic or hexagonal close
packed lattice ($K=12$, $p_\mathrm{c}=0.199$ in both cases)
\cite{Mar97}. It is close to the theoretical $p_\mathrm{c}$ of a
body-centered cubic lattice ($K=8$, $p_\mathrm{c}=0.246$). An
average coordination number of 8 seems reasonable in a typical
irregular structure.

It should be emphasized that two samples with an MgB$_2$ fraction
below 0.3 (but above 0.25) were apparently insulating. This could
be due to finite size effects, which are expected to result in a
larger percolation threshold, or, simply by the difficulty to make
contacts on the fragil spanning cluster, which could be even
disrupted during cooling after sintering due to mechanical
tensions induced by different thermal expansion coefficients of
MgO and MgB$_2$.

The two points corresponding to pure MgB$_2$
($p_\mathrm{MgB_2}>0.7$) were not taken into account for the fit
in either case, because they obviously deviate from the behaviour
of the mixed samples (Fig.~\ref{Figpc}). The added MgO powder
seems to reduce the average coordination number, possibly due to a
different morphology of grains, or by a reaction with MgB$_2$,
resulting in a new phase which blocks grain boundaries for
supercurrents. The non-zero resistivity of some samples is another
hint for the latter scenario. However, extrapolating
$p_\mathrm{c}^*$ of the pure MgB$_2$ samples to the perfectly
dense case (again with
$p_\mathrm{c}^*=p_\mathrm{c}/p_{\mathrm{MgB}_2}$, thus the values
of these two points are multiplied by about 0.7), one finds a
percolation threshold of about 0.19, which is low but not
unrealistic, since $p_\mathrm{c}$ was found to be even lower in
some systems \cite{Lee86,Gri06}.

\begin{figure} \centering \includegraphics[clip,width=0.5\textwidth]{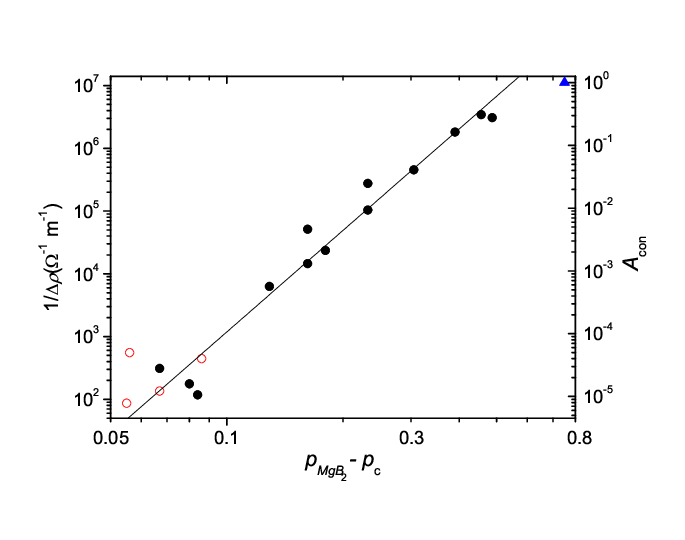}
\caption{Change of the conductivity with the MgB$_2$ volume
fraction. The line is a fit to the universal behaviour
(Equ.~\ref{Ap}) with $t=5.4$. The open symbols correspond to
samples with finite resistivity in the superconducting state. The
triangle in the upper right corner is the expected value for
perfect connectivity (9\,$\mu\Omega$cm).} \label{Figsigma}
\end{figure}

The phonon contribution to the resistivity was introduced as a
measure of the effective area in MgB$_2$ \cite{Row03}, since
$1/\Delta\rho$ should be proportional to the cross section over
which the current flows, at least if no other conducting phases
exist in the sample. $\Delta\rho$ is not expected to be influenced
by the observed changes of the superconducting properties.

Fig.~\ref{Figsigma} illustrates that the conductivity can be
nicely described by the universal behaviour (Equ.~\ref{Ap}). The
best approximating (only zero resistivity samples) transport
exponent, $t$, is 5.4, which is much higher than the theoretically
universal value of $\sim 2$ in three dimensional systems
\cite{Sta92}. The transport exponent in disordered continuum
systems, which can be described by the so called Swiss-cheese
model, is predicted to be only slightly higher, around 2.5
\cite{Hal85}. However, larger values were frequently observed
experimentally in real systems \cite{Lee86,Vio05,Gri06}. The
universal behaviour is predicted only for the vicinity of
$p_\mathrm{c}$ (which is not accessible to the present
experiment), thus the application to rather high fractions, p, is
problematic and might result in higher transport exponents. As
shown recently \cite{Joh08}, the transport exponent retains its
universal value only very close to $p_\mathrm{c}$, if tunnel
currents are taken into account and much higher apparent values
are expected at higher fractions $p$. One might speculate, that
connections, which are blocked by small MgO particles or layers,
might be transparent for tunneling currents leading to this high
transport exponent. This could also explain the finite resistivity
of some samples well below $T_\mathrm{c}$, if the local current
exceeds the maximum Josephson current.

Yamamoto et al. \cite{Yam07} found a larger percolation threshold
of 0.31, but a much weaker dependence of $\Delta\rho$ (and $J_c$)
on the volume fraction of MgB$_2$, which was varied between 0.44
and 0.87 by different preparation techniques (all samples were
prepared in-situ). Their non-superconducting volume consisted
mainly of pores. Thus, it is obvious that pores and MgO particles
have a different influence on the properties of MgB$_2$. This is
also supported by the present data, since extrapolation of the
data to $p_\mathrm{MgB_2}=1$ does not lead to the expected value
of $\Delta\rho\sim 9\,\mu\Omega$cm (triangle in
Fig.~\ref{Figsigma}), but to $1.67\,\mu\Omega$cm, although this
discrepancy could result also from a change in behaviour near
$p_\mathrm{MgB_2}=1$. However, we also found a smaller
extrapolated $p_\mathrm{c}$ in the pure MgB$_2$ samples (see
above).

\begin{figure} \centering \includegraphics[clip,width=0.5\textwidth]{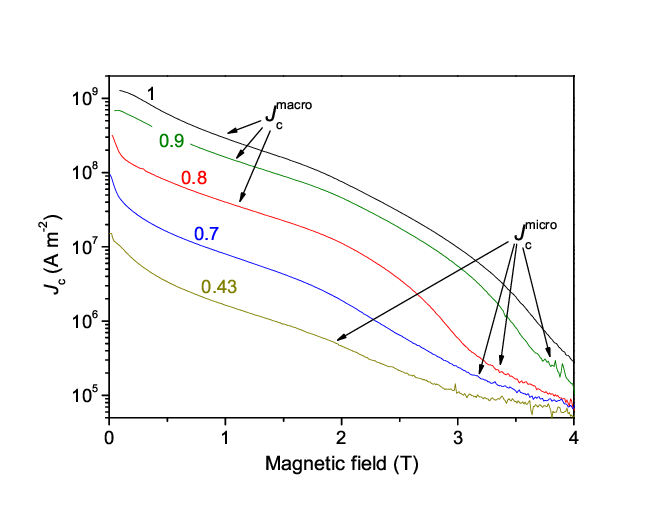}
\caption{Influence of the nominal MgB$_2$ content (labels) on the
critical current densities at 20 K.} \label{FigJc20K}
\end{figure}

The critical current densities at 20 K are presented in
Fig.~\ref{FigJc20K}. $J_c$ at self field is around
10$^9$\,Am$^{-2}$ in the pure sample, thus considerably smaller
than in typical in-situ prepared samples. $J_c$ in a sample of
comparable connectivity ($\Delta\rho$ was 29\,$\mu\Omega$cm in our
pure sample) was found to be around 4 times higher in
Ref.~\cite{Yam07}. This can be explained by the different grain
size. The grain size of in-situ prepared samples is generally
smaller (e.g. $\sim$ 100\,nm~\cite{Yam07}) than in ex-situ samples
(e.g. up to 1\,$\mu$m~\cite{Bir08}). In particular our long heat
treatment favors grain growth. Thus, the density of grain
boundaries, which are the dominant pinning centers in MgB$_2$, is
expected to be low in the present samples.

The critical current densities decrease with MgB$_2$ content. The
steep slope at higher fields (e.g. between 2 and 3 T in sample
0.8) gradually disappears. This strong decrease in $J_\mathrm{c}$
can be associated with the fast reduction in effective cross
section (for high currents). At higher fields, after the
decomposition of the spanning cluster, $J_c(B)$ becomes flatter
again, but the currents are now flowing on a different length
scale, namely on disconnected clusters of grains. Note that the
evaluation of $J_\mathrm{c}$ assumes the currents to flow
macroscopically around the sample. One obtains the corresponding
macroscopic current density $J_\mathrm{c}^\mathrm{macro}$, if this
condition is fulfilled. The evaluation becomes invalid if the
currents, which flow on a smaller length scale,
$J_\mathrm{c}^\mathrm{micro}$, generate a significant contribution
to the total magnetic moment \cite{Hor08}. These additional
currents lead to an overestimation of the macroscopic critical
current density, because they increase the measured moment. On the
other hand, $J_\mathrm{c}^\mathrm{micro}$ is underestimated, if
the macroscopic currents become small or even zero, because the
area enclosed by the microcurrents is much smaller than the sample
geometry and a priori unknown. Only if the macroscopic currents
dominate, the correct $J_\mathrm{c}^\mathrm{macro}$ is obtained,
in all other case the derived \emph{magnetic} $J_\mathrm{c}$ is
wrong. Nevertheless, this \emph{magnetic} $J_\mathrm{c}$
illustrates the differences between samples with and without large
macroscopic currents. Sample 0.43 was apparently insulating, thus
the spanning cluster is very weak or not existing at all, even at
zero field. Currents flowing on a microscopic length scale are
expected under all conditions (when superconducting grains exist),
but it depends on the field and the amount of MgB$_2$ whether they
give a negligible (small fields and large MgB$_2$ concentration),
important, dominant or even the only contribution (small MgB$_2$
concentration or large fields) to the signal. Their importance
generally increases with increasing field and decreasing MgB$_2$
concentration. The slope of $J_\mathrm{c}(B)$ is similar in all
samples up to about 2\,T (except in the low field region, where
geometry dependent self field effects become important), since the
length scale of current flow does not change
(Fig.~\ref{FigJc20K}). The decomposition of the spanning cluster,
if existing, always starts near $B_\mathrm{c2}/\gamma$. In
principle the weakening of the spanning cluster starts exactly at
$B_\mathrm{c2}/\gamma$, where the number of superconducting grains
$p$ starts to decrease, but if $p$ is assumed to depend on $J$
\cite{Eis03} ($p$ decreases with $J$) the ``spanning cluster for
high currents'' decomposes at smaller fields. After the
destruction of the spanning cluster, the average size and number
of the disconnected clusters further depend on
$p_\mathrm{MgB_2}p(B,J)$, leading to different slopes of the
\emph{magnetic} $J_\mathrm{c}$ at high fields. Note that
$B_\mathrm{\rho=0}$ derived from resistive measurements falls into
this region, e.g. 4.5\,T in sample 0.8. The influence of the
macroscopic currents in the remaining spanning cluster on the
magnetization is small, they might be even described by a network
of tunneling junctions \cite{Row03}.

\begin{figure} \centering \includegraphics[clip,width=0.5\textwidth]{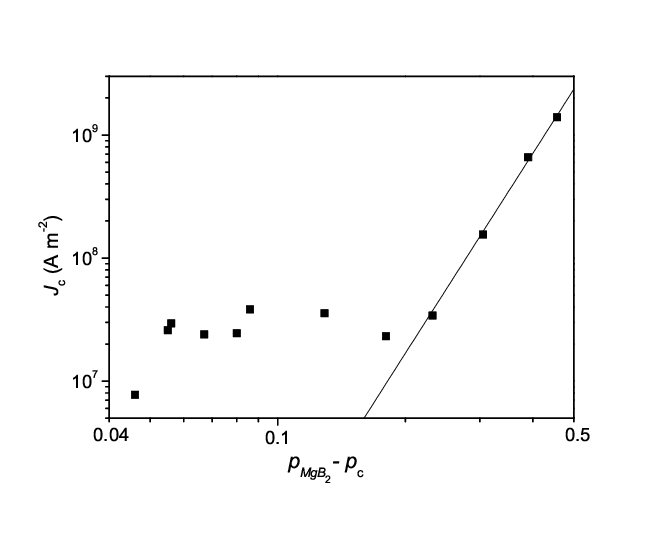}
\caption{Critical current densities at 5\,K and 0.5\,T.
$p_\mathrm{c}$ was assumed to be 0.25.} \label{FigJcp}
\end{figure}

The derived critical currents are a useful approximation for the
macroscopic currents at low MgO concentrations and low magnetic
fields. $J_\mathrm{c}(0.5\,\mathrm{T}, 5\,\mathrm{K})$ is plotted
as function of the MgB$_2$ volume fraction (reduced by
$p_\mathrm{c}=0.25$) in Fig.~\ref{FigJcp}. The transport exponent
$t=5.4$ is illustrated by the line graph and describes the data
very well, down to a volume fraction of about 0.5. The influence
of the microscopically circulating currents becomes dominant at
lower concentrations. The observation of the same reduction in
cross section for normal and supercurrents, which was also found
for porous samples \cite{Yam07}, supports the idea of using
$\Delta\rho$ to estimate the reduction in $J_\mathrm{c}$ due to a
reduced connectivity.

\section{Conclusions}

The percolative current transport was investigated in mixed
MgB$_2$/MgO samples. The susceptibility, the critical current, the
conductivity, and the superconducting transition width increase
with MgO content. The expected change in the relevant length scale
for shielding currents was observed in magnetization measurements.
An unusually high transport exponent of 5.4 and a percolation
threshold of about 0.25 were found. The latter is higher than in
porous MgB$_2$ ($p_\mathrm{c}<0.2$), which has a smaller
dependence of the transport properties on the superconducting
volume fraction than the mixed MgB$_2$/MgO samples. A reduction in
$J_\mathrm{c}$ by a factor of 2 can be expected by the presence of
only 10\,\% MgO.

The phonon contribution to the resistivity was confirmed to be a
useful measure for the decrease in $J_\mathrm{c}$ due to poor
connectivity or density.

\begin{acknowledgments}
The authors wish to thank Herbert Hartmann for technical
assistance.
\end{acknowledgments}


\end{document}